\documentclass[%
prl,twocolumn,%superscriptaddress,
 amsmath,amssymb,
 aps,
]{revtex4-2}

\usepackage{graphicx}% Include figure files
\usepackage{dcolumn}% Align table columns on decimal point
\usepackage{bm}% bold math
\usepackage{physics}
\usepackage{xcolor}
\usepackage{amsmath, amssymb}
\usepackage{braket}
\begin{document}

\preprint{APS/123-QED}

\title{Theory of a two-dimensional anharmonic piezoelectric crystal resonator}
\author{Karl H. Michel}
\email{karl.michel@uantwerpen.be}
\affiliation{Department of Physics \& NANOlab Center of Excellence, University of Antwerp, Groenenborgerlaan 171, B-2020 Antwerp, Belgium}

\author{Cem Sevik}
\email{cem.sevik@uantwerpen.be}
\affiliation{Department of Physics \& NANOlab Center of Excellence, University of Antwerp, Groenenborgerlaan 171, B-2020 Antwerp, Belgium}

\author{Milorad V. Milo\v{s}evi\'c}
\email{milorad.milosevic@uantwerpen.be}
\affiliation{Department of Physics \& NANOlab Center of Excellence, University of Antwerp, Groenenborgerlaan 171, B-2020 Antwerp, Belgium}

\date{\today}
\begin{abstract}
We developed a lattice dynamical theory of an atomically-thin compressional piezoelectric resonator. Acoustic and optical dynamic displacement response functions are derived and account for frequency-dependent electromechanical coupling. The dynamic susceptibilities for the direct and the converse piezoelectric effects are found equal. The mechanical resonant behavior of longitudinal in-plane displacement waves is investigated as a function of the lateral crystal size and of temperature in the classical and in the quantum regime. In the former case the quality factor of the resonator is inversely proportional to temperature and to crystal size. Below a cross-over temperature the quantum zero-point fluctuations become dominant and put an upper limit on the quality factor which is size independent. As experimentally relevant examples, the theory is applied on two-dimensional hexagonal boron nitride and molybdenum disulfide.
\end{abstract}

\maketitle
\newpage

%\section{\label{sec:int}Introduction}
The exploration of resonant mechanical motion in nano-electromechanical systems (NEMS), such as one-dimensional (1D) nanowires~\cite{reference1} and nanotubes~\cite{reference2} and two-dimensional (2D) crystals~\cite{reference3, reference4}, is a growing field of present day research~\cite{reference5} at the interface of atomic and solid-state physics. The ultra-light weight and high mechanical strength of NEMS devices offers new perspectives in fundamental physics, enabling inertial mass sensing at an atomic scale~\cite{reference6, reference7}, ultra-high force sensitivity~\cite{reference8, reference5}, and measurement quantum control of zero-point motion~\cite{reference9, reference10}. In addition, nanoscale resonators have a promising future in engineering, offering the integration of nanomechanics into the readily booming nanoelectronics~\cite{reference11}.  

Electromechanical coupling is at the origin of piezoelectricity~\cite{reference12}, the change of crystal polarization under applied stress, and its converse, the change of crystal shape under an electric field. Already in 1917, Langevin realized the importance of resonant properties in piezoelectric quartz crystals~\cite{reference13}. Nowadays, piezoelectric resonators have a plethora of applications, from quartz clocks to ultrasonic transducers in medical imaging. On a NEMS scale, subsequently to theoretical works on boron-nitride nanotubes~\cite{reference14, reference15}, mono and multilayer systems of hexagonal boron-nitride ($h$-BN)~\cite{reference16, reference17, reference18, reference19} and transition-metal dichalcogenides (TMDs)~\cite{reference18}, piezoelectricity has been observed in single and multilayer crystals of 2H-MoS$_2$ (molybdenum disulfide)~\cite{reference20,reference21}. However, 
dynamical resonant phenomena related to anharmonicities and quantum mechanics in 2D piezoelectric crystals have still to be explored.  

In this Letter, we develop an analytic lattice dynamical theory for an in-plane compressional 2D piezoelectric crystal resonator. To date, piezoelectric nanoelectromechanical resonators have predominantly been realized on the basis of aluminum nitride thin films~\cite{reference22}. There, as also in experimental work on layered atomically-thin resonators reviewed in Ref.~\onlinecite{reference5}, resonances of out-of-plane flexural modes have been investigated. In these membrane-like or drumhead resonators the vibrational restoring forces are due to pre-tension~\cite{reference4}. In contradistinction, the in-plane restoring forces of a compressional 2D crystal resonator are due to Young's modulus. Since the latter is stronger than the pre-tension, fundamental in-plane resonant modes have higher frequencies but smaller displacement fluctuations than drumhead modes. Hence the nonlinear dynamic characteristics~\cite{reference23} that require theoretical concepts pertaining to mesoscopic physics~\cite{reference24} need not be considered here.           

\begin{figure}[b]
\includegraphics[width=\linewidth]{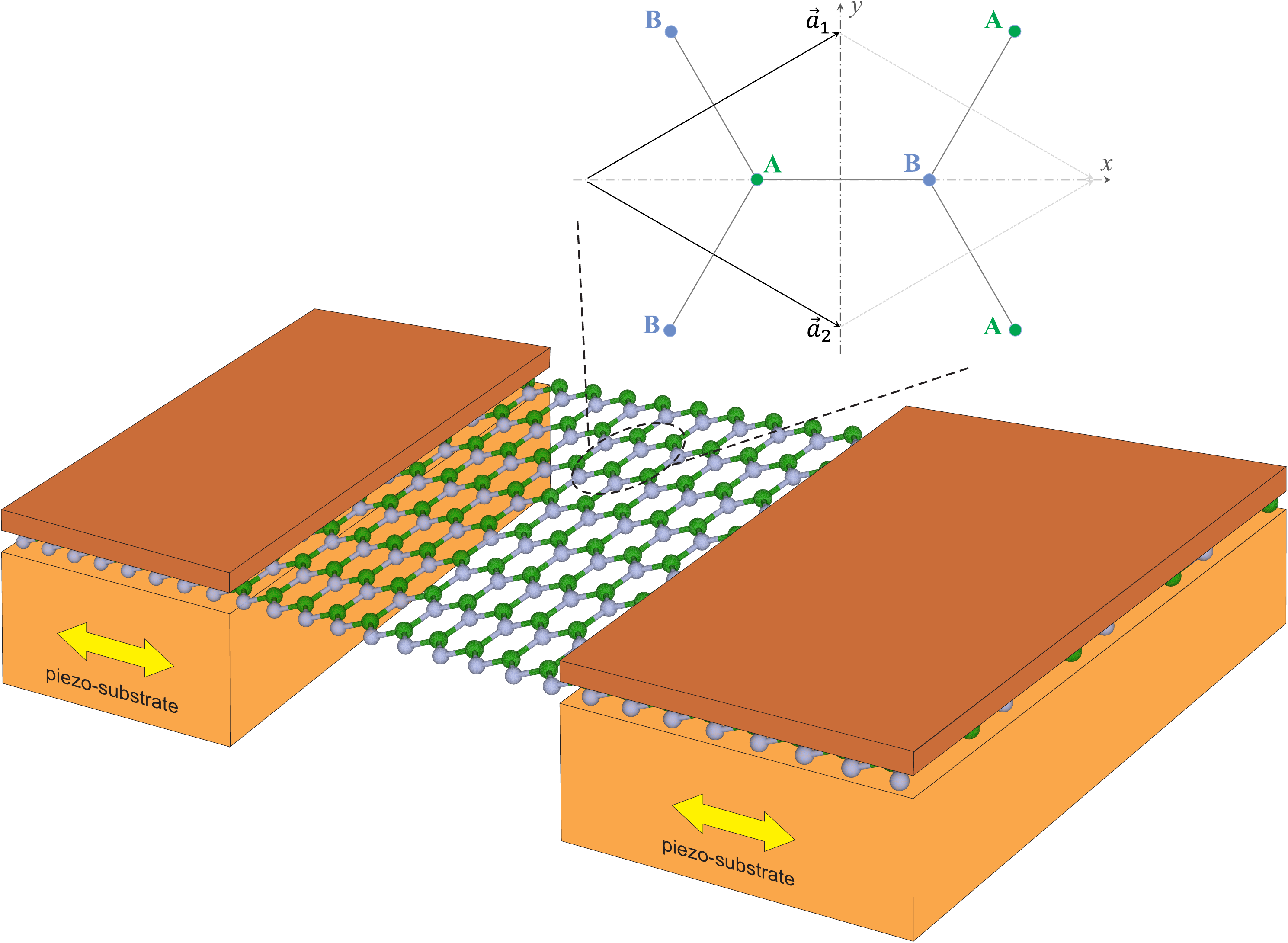}
\caption{Oblique view of an exemplified in-plane beam resonator based on a hexagonal 2D material, as studied in this Letter. Inset shows the unit cell of the 2D hexagonal crystal of point group symmetry D$_{3h}$, with two ions A and B, and their nearest neighbors ($|\vec{a_1}|$ = $|\vec{a_2}|$ = $a$).\label{figure-str}}
\end{figure}

We consider a model of a 2D hexagonal crystal of point group symmetry D$_{3h}$. The crystal of length $L$ and comparable width consists of $N$ unit cells, with two ions A and B per cell, as illustrated in Fig. \ref{figure-str} and Supplementary Materials File (SMF). The equilibrium positions are $\vec{X}(\vec{n}, \kappa)$ where $\vec{n}=\{n_1, n_2\}$ with $n_1$, $n_2$ integer numbers label the unit cells, and $\kappa \in \{\mathrm{A}, \mathrm{B}\}$. The particle masses are $m_A$ and $m_B$, and the effective ionic charges are $e_{\mathrm{A}}^{*}$ and $e_{\mathrm{B}}^{*}$, with $e_{\mathrm{A}}^{*} = -e_{\mathrm{B}}^{*}$. In case of 2D $h$-BN, A stands for nitrogen and B for boron, in case of 2H-MoS$_2$, A corresponds to the two sulfurs, taken to move in unison, while B stands for Mo.
Particle displacements away from equilibrium positions are denoted by vectors $u_i(\mathbf{n},\kappa)$, $i = \{x, y, z\}$.
Lattice waves are denoted by displacement fields $u_i^\kappa (\vec{q})$, where $\vec{q}$ is a wave vector in the hexagonal Brillouin zone (BZ). 
The dynamics of the crystal is described by a vibrational Hamiltonian~\cite{reference25} of the form $H=K+H_2+H_3$, where K stands for the kinetic energy, the harmonic term $H_2$ is quadratic and the anharmonic term $H_3$ cubic in the displacements $u_i^\kappa (\vec{q})$. 
Within linear response theory~\cite{reference26}, the expectation value $\Delta u_1^{\kappa}(\vec{q}, z)$ of longitudinal in-plane displacement waves induced by perturbation $H^{\prime}=-\sum_{\rho}u_1^{\rho}(-\vec{q})F_1(\vec{q},\Omega)/\sqrt{m_{\rho}}$, where $F_1(\vec{q},\Omega)$ is an external force of wave vector $\vec{q}=(q_1,0)$ and circular frequency $\Omega$, is given by
\begin{equation}
\Delta u_1^{\kappa}(\vec{q},z) =-\sum_{\rho}D_{11}^{\kappa\rho}(\vec{q},z)F_1^{\rho}(\vec{q}, \Omega)/\sqrt{m_{\rho}}.
\label{eq1}
\end{equation}
Here
\begin{equation}
D_{11}^{\kappa\rho}(\vec{q},z)=\braket{\braket{u_{1}^{\kappa}(\vec{q});u_{1}^{\rho}(-\vec{q})}}_{z},
\label{eq2}
\end{equation}
with $z=\Omega+i\varepsilon$, $\varepsilon\rightarrow +0$, is the Fourier component of frequency $\Omega$ of the time-dependent displacement-displacement correlation function (also called propagator or retarded thermal Green function~\cite{reference27}). Using field-theoretical methods~\cite{reference28}, we derive by means of the Hamiltonian $H$ the 2$\times$2 Dyson equation:
\begin{equation}
\left(\mathbb{I} z^2-\mathbf{D}(\vec{q})-\mathbf{\Sigma}_{11}(\vec{q},z)\right)\mathbf{D}_{11}(\vec{q},z)=\mathbb{I}.
\label{eq3}
\end{equation}
Here $\mathbb{I}$ is the unit matrix, $\mathbf{D}_{11}(\vec{q})$ is the dynamical matrix that accounts for the harmonic restoring forces and $\mathbf{\Sigma}_{11}(\vec{q},z)$ is the anharmonic phonon self-energy, also called polarization operator (see SMF). Using Born's long-wavelength expansion and transforming from the particle representation $\{\kappa,\rho\}$ to optical and acoustic displacement waves $u_1^{\xi}(\vec{q})$ and $u_1^{\eta}(\vec{q})$ respectively~\cite{reference16}, we rewrite Eq.~\eqref{eq3} as a system of coupled optical and acoustic in-plane propagator equations:  
\begin{eqnarray}
& &
\begin{pmatrix}
z^2-\omega_{E^{\prime}}^2 -\Sigma_{E^{\prime}}(z)  &  -iq_1\left(D_{11,1}^{\xi\eta}+\Sigma_{11,1}^{\xi\eta}(z)\right)\\
-iq_1\left(D_{11,1}^{\eta\xi}+\Sigma_{11,1}^{\eta\xi}(z)\right)  & \;\;  z^2-q_1^2\left(c_L^2+\Sigma_{11,11}^{\eta\eta}(z)\right)
\end{pmatrix}
\nonumber \\
& &
\times
\begin{pmatrix}
D_{E^{\prime}}^{\xi\xi}(z) & D_{11}^{\xi\eta}(\vec{q},z)\\
D_{11}^{\eta\xi}(\vec{q},z) & D_{11}^{\eta\eta}(\vec{q},z)
\end{pmatrix}=
\begin{pmatrix}
1 & 0\\
0 & 1
\end{pmatrix}.
\label{eq4}
\end{eqnarray}
Here, $\omega_{E^{\prime}}$
is the frequency of the doubly degenerate E$_{2g}$ mode~\cite{reference29, reference30}, $c_L$ the acoustic longitudinal displacement velocity, and $D_{11,1}^{\xi\eta}=-D_{11,1}^{\eta\xi}$ and $\Sigma_{11,1}^{\xi\eta}(z)=-\Sigma_{11,1}^{\eta\xi}(z)$ are static and dynamic opto-acoustic (viz. electromechanical) couplings. Solution of Eq.~\eqref{eq4} yields the opto-acoustic dynamic response function:
\begin{equation}
D_{11}^{\xi\eta}(\vec{q},z) = iq_1D_{E^{\prime}}^{\xi\xi}(z)\left(D_{11,1}^{\xi\eta}+\Sigma_{11,1}^{\xi\eta}(z)\right)D_{11}^{\eta\eta}(\vec{q},z),%=-D_{11}^{\eta\xi}(\vec{q},z),
\label{eq5}
\end{equation}
with $D_{E^{\prime}}^{\xi\xi}(z)= \left[z^2-\omega_{E^{\prime}}^2-\Sigma_{E^{\prime}}^{\xi\xi}(z)\right]^{-1}$ and $D_{11}^{\eta\eta}(\vec{q},z) =\left[z^2-q_1^2(c_{L}^2+\Sigma_{11,11}^{\eta\eta}(z))\right]^{-1}$.
The optical and acoustic displacements are related to the in-plane polarization $P_1(\vec{q})$ and inhomogeneous strain $\varepsilon_{11}(\vec{q})$ variables by 
\begin{equation}
P_{1}(\vec{q})=\frac{u_{1}^{\xi}(\vec{q})}{\sqrt{\mu}}\frac{e_{\mathrm{B}}^{*}}{A_{2D}}, %\;\; i \in \{1,2\}; \:\:\:\:\:\:\: 
\;\;\;\varepsilon_{11}(\vec{q})=\frac{iq_1u_{1}^{\eta}(\vec{q})}{\sqrt{m}},
\label{eq6}
\end{equation}
where $\mu$ and $m$ are the reduced and the total mass per unit cell, and $A_{2D}$ is the cell area. Within linear response theory we calculate the expectation value $\Delta P_1(\vec{q}, z)$ of the electrical polarization induced by a dynamic stress $\sigma_{11}(\vec{q},z)$ and obtain the dynamic susceptibility for the direct piezoelectric effect, as
\begin{equation}
\chi_{11}^{P\sigma}(\vec{q}, z) =iq_{1}e^{*}_{B}D_{11}^{\xi\eta}(\vec{q},z).%{\sqrt{\mu m}}
\label{eq7}
\end{equation}
Likewise, calculating the inhomogeneous frequency-dependent tensile strain $\Delta\varepsilon_{11}(\vec{q},z)$ induced by an electric field $E_1(\vec{q},\Omega)$, we obtain the dynamic susceptibility for the converse piezoelectric effect, as
\begin{equation}
\chi_{11}^{\varepsilon E}(\vec{q},z) = -iq_1e^{*}_{B}D_{11}^{\eta\xi}(\vec{q},z).
    \label{eq8}
\end{equation}
Using the symmetry property $D_{11}^{\eta\xi}(\vec{q},z)=-D_{11}^{\eta\xi}(-\vec{q},z)$, we find
\begin{equation}
\chi_{11}^{\varepsilon E}(\vec{q},z) = \chi_{11}^{P \sigma}(\vec{q},z),
\label{eq8-b}
\end{equation}
i.e. the longitudinal dynamic susceptibilities of the direct and the converse piezoelectric effects are equal. 

Irrespective of the nature of the external perturbation we now study the in-plane lengthwise resonant mechanical motion near the harmonic frequencies, $\Omega_n=c_Lq_1(n)$, with $q_1(n)=n\pi/L$, $n \in 1, 2, 3, ...$ and $L$ is the length of the sample. With $\Omega_n\ll\omega_{E^{\prime}}$ and $\Sigma_{11,1}^{\xi\eta}(z)\ll D_{11,1}^{\xi\eta}$, we may simplify the dynamic piezoelectric susceptibility to 
%Turning to the study of in-plane lengthwise resonant mechanical motion near the harmonic frequencies, $\Omega_n=c_Lq_1(n)$, where $q_1(n)=n\pi/L$, $n \in 1, 2, 3, ...$ and $L$ is the length of the sample, we may simplify the converse piezoelectric susceptibility to 
\begin{equation}
\chi_{11}^{\varepsilon E}(\vec{q}, z) =\frac{e^{*}_{B}q_{1}^{2}(n)D_{11,1}^{\xi\eta}}{\sqrt{\mu m}\omega_{E^{\prime}}^{2} \left\{z^2-q_1^2(n)[c_L^2+\Sigma_{11,11}^{\eta\eta}(z)]\right\}}.
\label{eq9}
\end{equation}
%Here we have neglected $\Omega^{2}_n+\Sigma^{\xi\xi}(z)$ in comparison with $\omega^2_{E^{\prime}}$ and $\Sigma_{11,1}^{\xi\eta}(z)\ll D_{11,1}^{\xi\eta}$.
Due to energy and wave-vector conservation, the anharmonic scattering processes in $\Sigma_{11,11}^{\eta\eta}(z)$ are dominated by the scattering of the resonant mode with pairs of low-lying out-of-plane acoustic (flexural) phonons of dispersion $\omega(k)=\sqrt{\kappa_0}k^2$, where $\kappa_0$ is the bending rigidity. We notice that this type of anharmonic interaction is at the origin of low temperature anomalies in 1D and 2D crystals~\cite{lifshitz1952thermal, lindsay2011flexural}. We proceed to obtain up to second order in the anharmonic force constants 
\begin{eqnarray}
&&\Sigma_{11,11}(z) = \nonumber\\ &&\hbar C_{11}^{\eta\eta} \int_{\omega_{inf}}^{\omega_{sup}}d\omega\left[1+2n(\omega)\right]\left[\frac{1}{z-2\omega}-\frac{1}{z+2\omega}\right],
\label{eq10}
\end{eqnarray}
where $\omega_{inf}=\sqrt{\kappa_0}(2\pi/L)^2$, $\omega_{sup}$ is determined by the boundaries of the Brillouin zone,
$C_{11}^{\eta\eta}$ is a positive material constant (see SMF), and $n(\omega)=[\exp(\hbar\omega/k_BT)-1]^{-1}$ is the ZA phonon distribution function at temperature $T$. 
The factor $[1+2n(\omega)]$ stems from the quantum-mechanical commutations, and the last two terms in Eq.~\eqref{eq10} account for absorption and emission of pairs of flexural phonons. We separate $\Sigma_{11,11}^{\eta\eta}(z)$ into real and imaginary parts as $ \Sigma_{11,11}^{\eta\eta}(z) = \Delta_{11,11}^{\eta\eta}(\Omega)+i\Gamma_{11,11}^{\eta\eta}(\Omega)$. Then, the real (reactive) and imaginary (absorptive) parts of the susceptibility read 
\begin{equation}
\chi_{11}^{\varepsilon E}(\vec{q},\Omega)^{^{\prime}}= \frac{q_1^2e^{*}_{B}D_{11,1}^{\xi\eta}\left[ \Omega^2-q_1^2 \left(c_L^2+\Delta_{11,11}^{\eta\eta}(\Omega)\right)\right]}{\sqrt{m\mu} \omega_{E^{\prime}}^2 |F_{11}(\Omega)|^2}, 
\label{eq18}
\end{equation}
and
\begin{equation}
\chi_{11}^{\varepsilon E}(\vec{q},\Omega)^{\prime\prime}= \frac{q_1^4e^{*}_{B}\Gamma_{11,11}^{\eta\eta}(\Omega)D_{11,1}^{\xi\eta}}{\sqrt{m\mu} \omega_{E^{\prime}}^2 |F_{11}(\Omega)|^2}, 
\label{eq19}
\end{equation}
where $F_{11}(\Omega) = \Omega^2-q_1^2\left(c_L^2+\Delta_{11,11}^{\eta\eta}(\Omega) + i \Gamma_{11,11}^{\eta\eta}(\Omega)\right)$. We infer that $\Delta_{11,11}^{\eta\eta}(\Omega)$ and $\Gamma_{11,11}^{\eta\eta}(\Omega)$ lead to resonance shifts and broadenings. 

We next discuss the resonance as a function of temperature and sample size. From Eq.~\eqref{eq10} we obtain
\begin{equation}
\Gamma_{11,11}^{\eta\eta} = -\pi\hbar C_{11}^{\eta\eta}\left[1+2n\left(\frac{\Omega_n}{2}\right)\right],
\label{eq192}
\end{equation}
where $n(\Omega_n/2)$ depends on temperature.

Near the absolute zero $n(\Omega_n/2)=\exp{-\hbar\Omega_n/2k_BT}\ll1$. At $T=0$, in the quantum-mechanical ground state (QGS), the finite value $\Gamma_{11,11}^{\eta\eta}=-\pi\hbar C_{11}^{\eta\eta}$ is caused by the anharmonic interactions of the in-plane phonons with the zero-point oscillations of the out-of-plane flexural modes, independent of $\Omega_n$ and hence independent of $L$. As is customary in the study of resonance~\cite{feynman1965flp}, we define a quality factor $Q_n=\Omega_n^2/[q_1^2(n)|\Gamma^{\eta\eta}(\Omega_n)|]$ as a measure of the energy stored in the resonator divided by the dissipated energy per cycle. It follows from the above discussion that in the QGS the zero-point fluctuations put an upper limit on the quality factor, namely
\begin{equation}
Q_{QGS} = \frac{c_{L}^{2}}{\pi\hbar C_{11}^{\eta\eta}},
\label{eq193}
\end{equation}
a value independent of length $L$ and harmonic number $n$.
We introduce the ratio, $Q_{QGS}/Q_n = 1+2n(\Omega_n/2)$ as the normalized energy dissipation (NED) with value 1 at $T=0$. We also define a crossover temperature $T_{QGS}$ by $2n(\Omega_n/2) = 1$, i.e. 
\begin{equation}
T_{QGS}=\frac{\hbar\Omega_n}{2k_B \ln3},
\label{eq19new}
\end{equation}
below of which the energy dissipation is dominated by the quantum zero-point motion rather than by thermal oscillations. Since $\Omega_n\sim n/L$, quantum fluctuations become relatively important at higher harmonics and small sample size $L$. One should notice that $T_{QGS}=\frac{1}{2}T_{QL}$, where $T_{QL}$ is the quantum-limited amplifier noise temperature due to zero-point fluctuations that effect the detection of gravitational radiation~\cite{caves1982quantum}. For quantum limited position measurements of a nanomechanical resonator, see Ref. \onlinecite{reference9}.  

In the classical regime at higher temperature, such that $2k_BT/\hbar\Omega_n\gg1$, and hence $|\Gamma_{11,11}^{\eta\eta}(\Omega_n)|=4\pi C_{11}^{\eta\eta}k_BT/\Omega_n$, the situation is reversed -- the quality factor is inversely proportional to temperature. We obtain the following scaling laws: for given $L$, $Q_n(T_1)/Q_n(T_2)\propto T_2/T_1$, for given $T$, $Q_n(L_1)/Q_n(L_2)\propto L_2/L_1$, and $Q_{n_1}(L)/Q_{n_2}(L)=n_1/n_2$.
\begin{table}[b]
\caption{\label{table1} Characteristic quantities of the compressional piezoelectric resonator, for different temperature $T$ and length $L$, at resonant circular frequency $\Omega_1$.} 
\begin{ruledtabular}
\begin{tabular}{lccclll}
 & T &$L$ & $\Delta_{11,11}^{\eta\eta}(\Omega_1)$ & $\Gamma_{11,11}^{\eta\eta}(\Omega_1)$ & $\Omega_1$ & $Q_1$ \\
 & K & $a$ & cm$^{2}$s$^{-2}$ & cm$^{2}$s$^{-2}$ & s$^{-1}$ &  \\\hline
$h$-BN & 0 & 10$^3$ & -1.04$\cdot$10$^9$ & -5.33$\cdot$10$^7$ & 2.63$\cdot$10$^{11}$ & 8.28$\cdot$10$^4$ \\
$h$-BN & 0 & 10$^5$ & -1.19$\cdot$10$^9$ & -5.33$\cdot$10$^7$ & 2.63$\cdot$10$^9$ & 8.28$\cdot$10$^4$ \\\hline
$h$-BN & 5 & 10$^3$ & -1.07$\cdot$10$^9$ & -5.33$\cdot$10$^8$ & 2.63$\cdot$10$^{11}$ & 8.27$\cdot$10$^3$ \\
$h$-BN & 5 & 10$^5$ & -1.29$\cdot$10$^9$ & -5.31$\cdot$10$^{10}$ & 2.63$\cdot$10$^9$ & 82.9 \\\hline
$h$-BN & 300 & 10$^3$ & -3.53$\cdot$10$^9$ & -3.19$\cdot$10$^{10}$ & 2.63$\cdot$10$^{11}$ & 138 \\
$h$-BN & 300 & 10$^5$ & -8.55$\cdot$10$^9$ & -3.19$\cdot$10$^{12}$ & 2.63$\cdot$10$^9$ & 1.38 \\\hline\hline
$2H$-MoS$_2$ & 0 & 10$^3$ & -5.65$\cdot$10$^8$ & -9.85$\cdot$10$^6$ & 7.14$\cdot$10$^{10}$ & 5.13$\cdot$10$^4$ \\
$2H$-MoS$_2$ & 0 & 10$^5$ & -5.94$\cdot$10$^8$ & -9.85$\cdot$10$^6$ & 7.14$\cdot$10$^8$ & 5.13$\cdot$10$^4$ \\\hline
$2H$-MoS$_2$ & 5 & 10$^3$ & -5.82$\cdot$10$^8$ & -3.63$\cdot$10$^8$ & 7.14$\cdot$10$^{10}$ & 1.39$\cdot$10$^3$ \\
$2H$-MoS$_2$ & 5 & 10$^5$ & -6.53$\cdot$10$^8$ & -3.63$\cdot$10$^{10}$ & 7.14$\cdot$10$^8$ & 13.9 \\\hline
$2H$-MoS$_2$ & 300 & 10$^3$ & -2.88$\cdot$10$^9$ & -2.17$\cdot$10$^{10}$ & 7.14$\cdot$10$^{10}$ & 23.2 \\
$2H$-MoS$_2$ & 300 & 10$^5$ & -5.71$\cdot$10$^9$ & -2.17$\cdot$10$^{12}$ & 7.14$\cdot$10$^8$ & 0.232 \\
\end{tabular}
\end{ruledtabular}
\end{table}

Using the material parameters specified in SMF, we have illustrated the theory by further numerical calculations for monolayer $h$-BN and 2H-MoS$_2$. The results in Table~\ref{table1} corroborate the above analytical scaling laws from the theory. In addition, under similar conditions of temperature and sample size, 2D $h$-BN is found to exhibit a higher resonance frequency and a higher quality factor than 2H-MoS$_2$, which is a consequence of the stronger restoring forces in the former material, also manifested in the phonon dispersions~\cite{reference29, reference30}. Likewise, at fundamental resonant frequencies and sample size $L=10^3a$, we obtain $T_{QGS}$ as 0.915~K and 0.245~K for 2D $h$-BN and 2H-MoS$_2$, respectively. 

\begin{figure}[t]
\includegraphics[width=\linewidth]{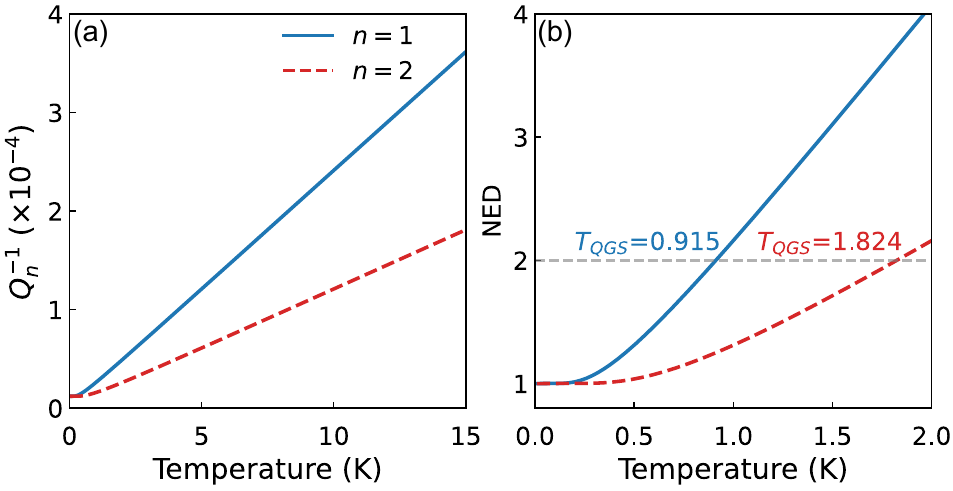}
\caption{The calculated $Q^{-1}_{n}$ and NED as a function of temperature for 2D $h$-BN of length $L=10^3a$. \label{figure2}}
\end{figure}

Experiments on out-of-plane monolayer TMD resonators~\cite{reference32} at Helium temperature have revealed quality factors up to 47$\times$10$^3$ at resonant frequency $\Omega_1\approx 57.6$~MHz. Likewise, the results of Table \ref{table1} show significant increase of the quality factor with cryogenic cooling, with maximum values reaching $Q_{QGS}$ at $T\rightarrow0$. In Fig.~\ref{figure2} we have plotted $Q^{-1}_{n}$, $n=1, 2$ (panel (a)) and the normalized energy dissipation (NED) $Q_{QGS}/Q_n$ (panel (b)) for a $h$-BN monolayer of $L=10^3a$ as a function of temperature. %The finite value 1 for NED at $T=0$ as a consequence of zero-point oscillations.
One notices from our preceding analysis and Fig.~\ref{figure2} that $T_{QGS}$ and thereby the range of the quantum regime, as well as the quality factor for $T\neq 0$, increase with harmonic number $n$.

\begin{figure}[t]
\includegraphics[width=\linewidth]{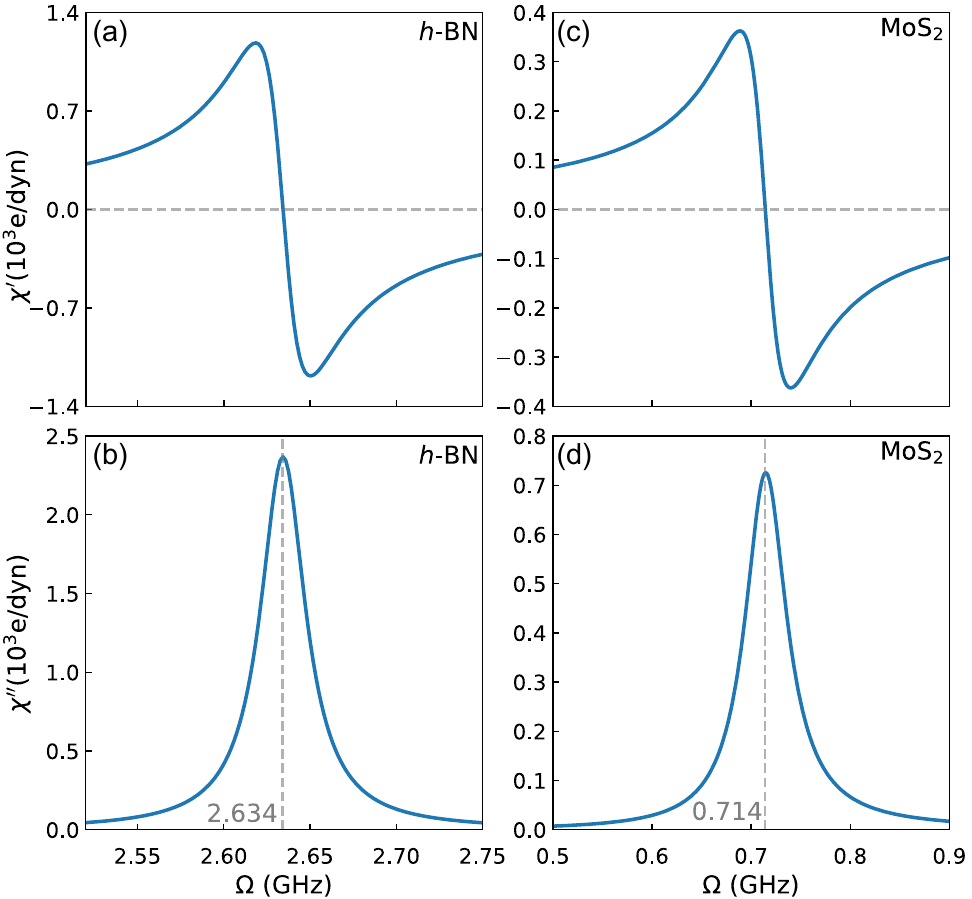}
\caption{Reactive and absorptive susceptibilities of 2D $h$-BN (a,b) and $2H$-MoS$_2$ (c,d) at $T=5$~K, each of length $L=10^5a$, near their respective resonant frequencies $\Omega_1=2.634 $~GHz and $714$~MHz, with respective quality factors $Q=82.9$ and 13.9.\label{figure3}}
\end{figure}

In Fig.~\ref{figure3}(a,c) we have plotted the reactive and absorptive susceptibilities $\chi^{\prime}(\Omega)$ and $\chi^{\prime\prime}(\Omega)$ of 2D $h$-BN for $L=10^5a\approx 25~\mu$m at $T=5$~K, which is in the midst of the classical regime. Here $T_{QGS}=9.15$~mK since the temperature range of the quantum regime is inversely proportional to the sample size. Plots of 2H-MoS$_2$ susceptibilites under the same conditions [cf. Fig.~\ref{figure3}(b,d)] show broader resonances near $\Omega_1=714$ MHz, with $Q=13.9$ (see also Table \ref{table1}).

Here a brief comparison with earlier published work is in order. Comparison with $Q^{-1}$ data measured in Ref.~\onlinecite{reference32} as a function of temperature and laser power suggests that there too the mechanical damping at low temperature is due to anharmonic phonon processes, while the magnitude of the $Q_n$ is comparable with the one obtained in the present work, see Fig.~\ref{figure2}(a). 

At room temperature, for a single layer MoS$_2$ compressional resonator of $L=10^4 a\approx 3~\mu$m, we obtain the resonant frequency $\Omega_1=1$~GHz and the quality factor $Q_1=2.32$. The experimental values~\cite{reference4} for a single layer MoS$_2$ drumhead resonator of comparable size are $\Omega_1\approx20$~MHz and $Q_1\approx40$. We attribute the difference in $\Omega_1$ values to the different restoring forces. Namely, the restoring forces are essentially determined by the Young modulus $Y_{2D}=122$~N/m in case of the compressional MoS$_2$ resonator and by the pre-tension $R=0.015$~N/m in case of the drumhead resonator~\cite{reference4}. On the other hand, the quality factor of the compressional resonator is much smaller than the one of the drumhead resonator at room temperature, which is a consequence of the large energy dissipation rate due to scattering of the in-plane resonant mode with out-of-plane thermal phonons. In contradistinction, the out-of-plane resonant vibrations of the drumhead resonator are due to a coherent motion of flexural modes with weak friction by the bath~\cite{reference24}. Similar reasoning should also apply to 2D $h$-BN, however we are not aware of such single-layer resonator experiments to date (for a resonator with thickness above 20 $h$-BN layers, see Ref.~\onlinecite{reference31}). 

We expect the present work to stimulate the experimental realization of a single-layer compressional resonators, with many potential applications in both fundamental physics~\cite{reference5} and nanoelectronics~\cite{reference11}. For example, when inserted in electrical circuits, such devices could function as crystal frequency controllers and stabilizers. For high precision experiments, besides the high resonant frequencies, a large quality factor $Q$ is required. In strained nanoscopic resonators, the needed enhancement of quality factors is realized by the mechanism of dissipation dilution \cite{reference33,reference34}. In the present case we infer that dissipation dilution, reducing $\Gamma^{\eta\eta}_{11,11}$, can be achieved by static tensile strain, thereby decreasing the anharmonic interaction and increasing the bending rigidity $\kappa_0$ (see $C^{\eta\eta}_{11}$, SMF). In any case, based on our results in Table~\ref{table1}, at cryogenic temperature we expect the in-plane piezoelectric resonator to typically reach resonant frequencies in the GHz range with a quality factor
$Q_1\approx10^3$. In the quantum regime, below $T_{QGS}$, $\Omega_1\approx10$~GHz with $Q_1\approx10^4$ should be accessible.

This work was supported by the Research Foundation-Flanders (FWO-Vlaanderen).

\bibliography{bibliography.bib}
\end{document}